# Speaker Identification using Speech Recognition

Syeda Rabia Arshad
Department of Computing (DoC),
School of Electrical Engineering
and Computer Science (SEECS)
National University of Sciences and
Technology (NUST)
Islamabad, Pakistan

Syed Mujtaba Haider
Department of Computer Science
School of Engineering
and Applied Sciences (SEAS)
Bahria University
Islamabad, Pakistan

Abdul Basit Mughal
Department of Computer Science
School of Engineering
and Applied Sciences (SEAS)
Bahria University
Islamabad, Pakistan

*Abstract*— The audio data is increasing day by day throughout the globe with the increase of telephonic conversations, video conferences and voice messages. This research provides a mechanism for identifying a speaker in an audio file, based on the human voice's biometric features like pitch, amplitude, frequency etc. We proposed an unsupervised learning model where the model can learn speech representation with limited dataset. Librispeech dataset was used in this research and we were able to achieve word error rate of 1.8.

## I. INTRODUCTION

Speaker identification is the process of identifying a person from their voice. Given a handful of voice samples, associating voices with their respective speakers using speech recognition requires identification of a person through their pitch, amplitude and other voice characteristics. A speaker's voice has some personal traits of that speaker like their accent, rhythm, unique vocal tract shape etc. Therefore, it is possible to identify a person's voice through the computer automatically, just like how humans are able to do it. These voice samples can be recorded through microphone or can be telephonic conversations. Irrespective of the language, speaker identification can work on any type of human voice. However, some letters or words are not present in all of the languages, but this research can still hold because the speech signals of the voice remain the same, regardless of the language the person is speaking. Speaker recognition has many real-world applications including biometric door unlocking in security-critical places, voice-based authentication of personal smart devices like laptops, cell phones etc., in transaction security of remote paying and bank trading, in forensics to investigate if a person is suspect to be guilty of a crime, or surveillance and automatic identity tagging.

Some of the main aspects of automatic speech recognition are speaker identification, verification and diarization. Speaker identification deals with simply associating speaker with their respective voice biometrics, speaker verification is the process of authenticating the voice when a speaker claims to be some specific person, and speaker diarization is the process of partitioning or labelling the audio segments by each speaker. This paper will mainly focus on speaker identification, and in some cases, speaker verification as well. This research can later on serve as a basis for multi-speaker identification as well, where we identify multiple speakers in an audio.

In this paper, we are going to implement some deep learning models to implement speaker identification task. Deep learning has some major advantages over other conventional methods, like its representation ability through which it is able to create highly abstract embedding features out of utterances, which helps in feature extraction from the voice samples.

## II. LITERATURE REVIEW

### A. Wav2Vec

Wav2Vec is a recently developed speech recognition method that masks the input of speech and solves a contrastive task defined over a quantization of the latent representations which are jointly learned. This method has been extended using two approaches i-e the self-supervised approach and the unsupervised approach.

Deep learning models often require huge amount of training data. But in many cases, data is unlabeled. Labeled data is much harder to come across than the unlabeled data. Which sometimes make it difficult to train the neural network as the training mainly rely on quality and quantity of data. This sometimes result in reduced accuracy of the model. Self-supervised learning has a paradigm to learn some general data representations from unlabeled examples of data and to fine-tune the model to labeled data. This technique has showed great success in Natural Language Processing and is also showing some promising results when applied to Computer Vision. The wave2vec 2.0 is the self-supervised framework for learning representation from raw audio encodes the speech audio through a multi-layer convolutional neural network which then further masks the spans of resulting latent speech representations. This is somewhat similar to masked language modeling where the task is to identify language. The latent representations are then fed to a transformer network for building contextualized representations and then model is trained via a contrastive task where the actual latent is to be distinguished from distractors. During training, the discrete units are learnt via gumbel softmax for representing the latent representations in contrastive tasks and after pre-training on the unlabeled data, the model is fine-tuned on the labeled data with a Connectionist Temporal Classification (CTC) loss.[1]

The other approach of wav2vec is with unsupervised learning. Which is often known as wav2vec-U, short for wav2vec unsupervised. Most of the current speech recognition systems require labeled training data which limits the technology to the languages spoken around the globe. It is hard to obtain feature-rich, labelled training dataset for each of the languages. As the name suggests, wav2vec-U is the method to train speech recognition systems without any labeled data. This framework leverages the self-supervised learning representations from wav2vec 2.0 to embed the speech audio and to partition the audio into the units by applying k-mean clustering method. In this method, first the simple representations of audio are learned with wave2vec 2.0 on unlabeled audio, clusters in the representations are identified with simple k-means clustering method, next segment representations by mean pooling the wav2vec representations are built. Now this is fed as an input to the generator which outputs a phenome sequence. This is sent as an input to the discriminator, which is similar to the unlabeled text, for performing adversarial training.[2]

The experimental results show the viability of this model for a number of settings and languages.

### B. Iterative Psuedo-Labeling (IPL)

Pseudo-Labelling recently has shown some great advancement in end-to-end speech recognition. Recent research has shifted to self- and semi-supervised learning algorithms in order to better utilize the effectiveness of unlabeled data. Iterative Pseudo-labeling is a semi-supervised learning algorithm which performs iterations of pseudo-labelling on unlabeled data. It fine-tuned the existing model at each iteration using a combination of both labeled and a proportion of unlabeled data. The main components of IPL are data augmentation and decoding with a language model. This is a simple and straightforward technique that can further scale on to large unlabeled datasets and it can show boost in performance in low resource environments.[3]

### C. Cross-lingual representation learning

Cross-lingual representation aims to learn representations generated from other languages to gain an improvement in performance. Unsupervised or pretrained representation learning does not require labeled data. It can learn from word embedding from unlabeled data as well. In this work, fine-tune of the transformer part is done instead of the parameters.[4]

## III. METHODOLOGY

### A. The Model

For extracting the representations from the audio, we first setup a multi-layered convolutional network that takes raw audio as input and extracts latent speech representations from it which is then fed to the transformer to extract information from the whole sequence rather than for each time stamp. For creating a network that builds representation over contiguous speech representations, we created a model that consists of encoder – transformer – quantization module.

The feature encoder consists of multiple blocks that contains a temporal convolutional network which is followed by layer normalization and GELU activation function. The audio which is fed as input to the encoder is normalized by using zero mean and unit variance. Since right now we are encoding the representations on each time stamp, the stride shift of the convolutional layer of our encoder determines how many time steps we are moving forward which will be sent as input to the transformer.

Now for building the context representations of the transformer, the output of the encoder convolutional network is fed to a context network which is combined with the transformer architecture. Instead of fixed positional embeddings which encodes absolute positional information, this model use a convolutional layer similar to which acts as relative positional embedding. We add the output of the convolution followed by a GELU to the inputs and then apply normalization.

The quantization module discretises the output of the feature encoder to a finite set of speech representations via product quantization for self-supervised training.

### B. The Dataset

The dataset that we are using is Librispeech dataset which is a large corpus of almost 1000 hours of recording in English speech. Our model learns the context representation of each time step and self-attention module extracts the dependencies of the entire sequence end-to-end.

### C. Training the model

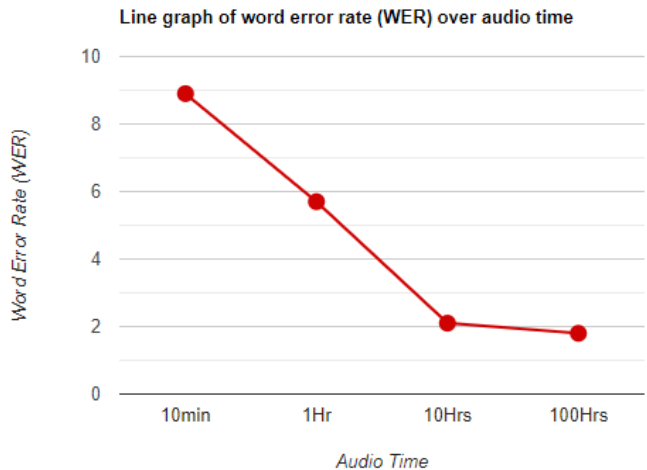

*Figure 1: Comparison of Word Error Rate (WER) over audio timeperiod*

To train the model we masked a little proportion of time steps in the feature of latent encoder space. We masked a proportion of the encoder outputs, or time steps before feeding them as input to the context network and substitute them with a pre-trained feature vector shared between all masked time steps. Then do not mask the inputs to the quantization module. To mask the latent representations output by the feature encoder, we randomly sample without the replacement a certain proportion of all time steps to be starting indices and then mask the subsequent M consecutive time steps from every sampled index; spans may overlap.

During the training, we are learning representations of speech audio by solving a contrastive task which is required to identify the true quantized latent speech representation for an already masked time step within a set of distractors. This is added to a codebook diversity loss to enable the model to use the codebook entries equally often. Now we move on to calculate the contrastive loss of our model. Given context network output, centred over the masked time step, the model needs to recognize the actual quantized speech representation in a set of quantized representations which includes t time-steps and K distractors. Distractors are equally sampled from other masked time steps of the same utterance.

Models are optimized by minimizing a contrastive loss and we mask the time-steps and channels during training which delays overfitting and significantly improves the final error, especially on the Libri-light subsets with a few labelled examples. After training we fine-tune the learned representations on labelled data and add a randomly initialized output layer on top of the Transformer to predict characters (Librispeech /Libri-light). The optimizers used in this research work were the Adam optimizer and a tri-state rate schedule where the learning rate is updated for the intial 10% of the updates, kept constant for the next of 40% and then gradually decomposition for the remainder.

The language model used for this work was n-gram language model where the value of n is 4. The reason for choosing 4-gram language model is that the data that our model has been trained on is very large. Apart from the pre-training, the fine tuning of the data is done on a huge corpus of LibriSpeech dataset, so for capturing most of the words in n-gram, we have to increase the value of n. So, it should be more than bigrams of trigrams.

## IV. RESULTS

The automated speech recognition system is evaluated by Word Error Rate (WER) where the overall error to recognize words is considered to be a fault of the ASR (Automated Speech Recognition) system. The model is tested on the 10 minutes of LibriSpeech LV-60k dataset. The model achieved the word error rate (WER) of 18. This result was achieved through our approach on Librispeech benchmark.

Moreover, we also compared the performance of previous related works in Automated Speech Recognition (ASR) systems which were supervised, semi-supervised and unsupervised.

**Table 1:** Comparison of models and their word error rates(WER)

| Model | Word Error Rate (WER) |
|---|---|
| Supervised | 2.1 |
| Semi-Supervised | 2.0 |
| Unsupervised (This Research) | 1.8 |

## V. ANALYSIS

The automated speech recognition model proposed in this research work performs good on the unsupervised data. However, some of the pre-training of the model is done on the labelled data in a supervised / semi-supervised fashion. The Librispeech dataset contains 10 min audio recordings to up-to 100hrs of audio recordings which helped in pre-training the model. Overall, this dataset was good for our task but we need more precise dataset with specific persons labelled along with their audio file. If we have more instances of specific people, labelled with their names then it will be helpful in the fine-tuning of our model. In this way, the model will be better able to recognize the speaker if their sample voice audio is played.

## VI. CONCLUSION AND FUTURE WORK

In this work, we proposed an automated speech recognition model to identify the speaker based on their biometric features of voice like pitch, amplitude etc. Since this model focuses on the audio features, this model is independent of the language a person is speaking and can identify the speaker regardless of the language they are speaking. Our model achieved the Word Error Rate (WER) of 1.8.

In the future, we would want to enhance the features of this model to multi-speaker identification where the model can identify each speaker from and audio file where more than one people are speaking. And it will be able to tell the time duration from an audio file where the person's voice is detected.